# Confinement-Induced One-Dimensional Magnetism in CrSBr Chains via Carbon Nanotube Encapsulation

*Diego López-Alcalá[1], Alberto M. Ruiz[1], Andrei Shumilin[1] and José J. Baldoví[1,*]*

[1]Instituto de Ciencia Molecular, Universitat de València, Catedrático José Beltrán 2, 46980 Paterna, Spain.

e-mail: j.jaime.baldovi@uv.es

**ABSTRACT**

Encapsulating low-dimensional magnetic materials within carbon nanotubes (CNTs) offers a compelling route to stabilize unconventional magnetic states and engineer quantum functionalities at the limit of miniaturization. In this work, we systematically investigate the structural, electronic, and magnetic properties of one-dimensional (1D) CrSBr chains encapsulated within CNTs using density functional theory (DFT) and spin dynamics simulations. We demonstrate the structural stability of CrSBr@CNT, where confinement and charge transfer cooperate to stabilize ferromagnetism in the 1D limit, which persists up to 50 K. These findings position CrSBr@CNT as a model platform for realizing 1D magnetism and establish CNT encapsulation as a powerful strategy for exploring emergent quantum spin phenomena and engineering nanoscale spintronic devices.

**INTRODUCTION**

The discovery of intrinsic magnetism in two-dimensional (2D) van der Waals (vdW) materials[1,2] has opened new frontiers in spintronics[3] and magnonics,[4] where low dimensionality and quantum confinement enable emergent spin phenomena and device functionalities.[5] These 2D magnets tunable magnetic order with long-range spin transport, and serve as versatile building blocks for assembling vdW heterostructures without chemical constraints.[6,7] A natural extension of this paradigm is the pursuit of magnetism in the one-dimensional (1D) limit, where the confinement of magnetic interactions to a single direction allows the emergence of exotic phenomena.

1D magnetic systems have already revealed signatures of quantum ballistic transport, anisotropic magnetoresistance, and even the emergence of Majorana fermions.[8–10] However, their synthesis remains challenging due to strong structural instabilities and curling, which hinder experimental access to their intrinsic properties.[11–14] In this context, carbon nanotubes (CNTs) emerge as exceptional host structures, providing mechanical stability, chemical inertness and an electrostatic environment conducive to the confinement and protection of 1D chains.[15,16] Encapsulation via chemical vapor transport (CVT) has proven especially effective,[17] enabling the realization of robust CNT-based heterostructures where confined 1D phases exhibit unique electronic,[18–20] catalytic[21,22] and magnetic properties,[23–27] which are absent in their bulk or 2D counterparts.[28–35]

Recent experimental efforts have demonstrated that single chains of $CrCl_3$ can be confined within CNTs,[27] exhibiting a spin-glass ground state at 3 K,[23] while encapsulated 1D β-W and $V_xTe_y$ chains display tunable magnetic configurations.[25,26] These advances underscore the potential of CNTs to stabilize 1D magnetism. Yet, the number of magnetic systems successfully encapsulated remains limited. Extending this strategy to other magnetic vdW semiconductors with intrinsic anisotropy and scalable synthetic routes remains an open challenge. Among these candidates, CrSBr stands out due to its strong in-plane anisotropy, semiconducting character, and relatively high Curie temperature in the monolayer ($T_C$ = 146 K).[36–38] Its structural anisotropy, stemming from distinct Cr–S–Cr and Cr–Br–Cr bond geometries, favors elongated growth along $a$ axis and promotes anisotropic spin and charge transport. Importantly, bulk CrSBr crystals can be readily synthesized via CVT,[39] making them accessible for controlled exfoliation, intercalation, and confinement.

In this work, we investigate the emergence of 1D magnetism in CrSBr nanoribbons (NRs) encapsulated within CNTs using first-principles calculations and atomistic spin simulations. We demonstrate that CNT confinement, combined with charge transfer at the interface, stabilizes robust ferromagnetic order in the 1D limit, while preserving the intrinsic anisotropic features of CrSBr. Our results reveal enhanced exchange couplings and coherent magnon propagation, establishing CrSBr@CNT as a prototypical platform for studying confined spin phenomena and for engineering 1D architectures for prospective applications in cutting-edge spin technologies.

**RESULTS**

CrSBr is a layered van der Waals (vdW) semiconductor with A-type antiferromagnetic (AFM) ordering and pronounced structural, electronic, and magnetic anisotropy. This anisotropy originates from the directional disparity in Cr–X–Cr bonding geometries: along the $a$-axis, Cr–S–Cr and Cr–Br–Cr angles are close to 90°, whereas along the $b$-axis, Cr–S–Cr angles approach 160°. As a result, the lattice exhibits a strong preference for uniaxial elongation along the $a$ direction, i.e. a feature that naturally predisposes CrSBr to form extended 1D architectures. This structural anisotropy has been shown to manifest in axis-dependent carrier transport and spin interactions[38], and makes CrSBr a promising platform for dimensional reduction toward 1D magnetism. In parallel, carbon nanotubes (CNTs) have emerged as versatile hosts for encapsulating low-dimensional materials, with experimentally accessible inner diameters ranging from 10 – 38 Å[28]. This size range makes them ideal candidates to accommodate a 1D NR of the magnetic semiconductor CrSBr with a width of 3 to 15 Å (Figure 1a and b).

To evaluate the stability of 1D CrSBr NRs with respect to their 2D counterparts, we perform density functional theory (DFT) calculations for three different widths of the NRs –2.4, 7.8 and 13.0 Å– upon electrostatic doping, revealing that 1D NRs along *a* direction of CrSBr can be stabilized when the system has an excess of electrons (Figure 1c). Notably, the required electron doping depends sensitively on the ribbon width, since wider ribbons retain characteristics of the 2D parent lattice and require additional doping to reach thermodynamic stability. Based on these results, we focus on a CrSBr 1D NR of 13 Å width for encapsulation into a CNT structure. As hosts, we consider CNTs with diameters ranging from 17.6 to 23 Å, providing sufficient space for hybridization and electron doping for the NR stabilization. With these considerations, we investigate: (i) the use of CNTs with diameters that have been experimentally employed to fabricate similar 1D hybrid nanostructures; (ii) the encapsulation of a sufficiently wide CrSBr 1D chain that can potentially be stabilized; (iii) the effect of CNTs inner cavity on the confinement of 1D nanomaterials.

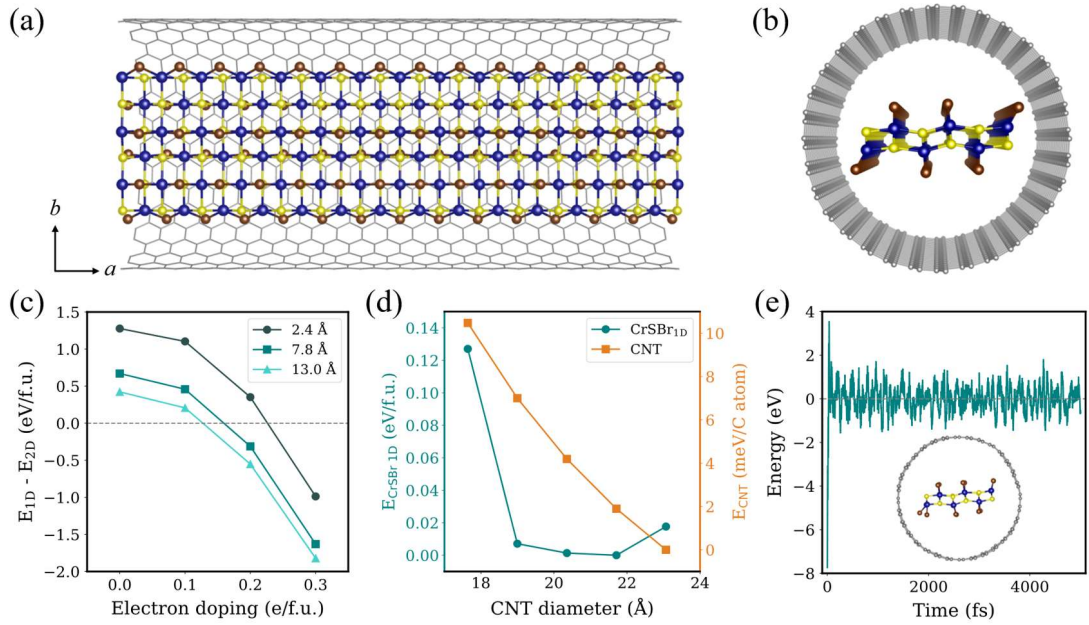

**Figure 1.** Top (a) and side (b) views of a CrSBr@CNT heterostructure. (c) Stabilization of 1D CrSBr NR with different widths upon electron doping. (d) Total energy of CrSBr NR relaxed inside CNTs of varying diameters (left) and energy per C atom for the corresponding CNTs (right). (e) AIMD simulation at 300 K in CrSBr@CNT with energies relative to the mean value. Inset image show a random snapshot of the structure.

Consequently, we fully relax the CrSBr@CNT structures described above. Figures 1a and b show the structure of the hybrid system, where CrSBr is only slightly distorted by the confinement, as it preserves its axis-dependent bonding angle. According to our results, the terminal Br atoms tend to approach the carbon skeleton due to the orbital hybridization that emerges at the interface. We compute the total energies of a CrSBr NR encapsulated in CNTs of different diameters for comparison (Figure 1d), revealing a noticeable decrease in energy for structures with CNT diameters larger than 18 Å. This behavior arises from the limited space in smaller CNTs, which induces significant distortions in CrSBr and leads to unfavorable conformations (Figure S1 and Table S1). Among these, the structure relaxed inside the armchair (16,16) CNT (21.7 Å diameter) shows the lowest relative energy, likely due to a favorable balance between space filling and hybridization with the carbon skeleton. Therefore, we use this structure as the reference for the following analysis. To examine the structural stability of CrSBr@CNT, we perform *ab initio* molecular dynamics (AIMD) simulations (see Methods). Room-temperature simulations (Figure 1e) show that the energy stabilizes, and structural snapshot confirm that the integrity of the system is preserved, supporting the thermodynamically stability of the proposed hybrid CrSBr@CNT structure.

Next, we explore the electronic structure of the hybrid system, as hybridization gives rise to emerging phenomena that significantly alter the electronic and magnetic properties of the heterostructure components. Electronic structure calculations for a 2D monolayer of CrSBr (Figure 2a) reveal a direct band gap at $\Gamma$, consistent with a semiconducting ground state. Highly dispersive bands are observed along the $\Gamma$–Y direction (*b* axis), while they flatten along the $\Gamma$–X direction (*a* axis), thus confirming the origin of the pronounced electronic anisotropy in CrSBr. Accordingly, we compute the band structure of the 1D CrSBr NR along *a* axis (Figure S2) and find that the characteristic electronic dispersion along the $\Gamma$–X direction and the semiconducting ground state of the 2D counterpart are preserved. This underscores that the pronounced electronic anisotropy of CrSBr remains robust under dimensional reduction. On the other hand, we simulate the band structures of the different armchair CNTs (Figure 2b and Figure S3), which reveal a metallic ground state characterized by high electron mobility. This is consistent with the extended $\pi$-electron system of the carbon skeleton. Figure 2c shows the component-resolved band structure of the CrSBr@CNT heterostructure. One can observe that noticeable charge transfer at the interface shifts the CrSBr (CNT) bands downwards (upwards). Importantly, the overall band structure remains largely preserved compared to that of the isolated constituents. Figure 2c presents the projected density of states (PDOS) of CrSBr@CNT, highlighting that spin polarization occurs exclusively on CrSBr, while CNT states remain unpolarized. These observations indicate modest interaction between the heterostructure components, which helps maintain their intrinsic electronic properties.

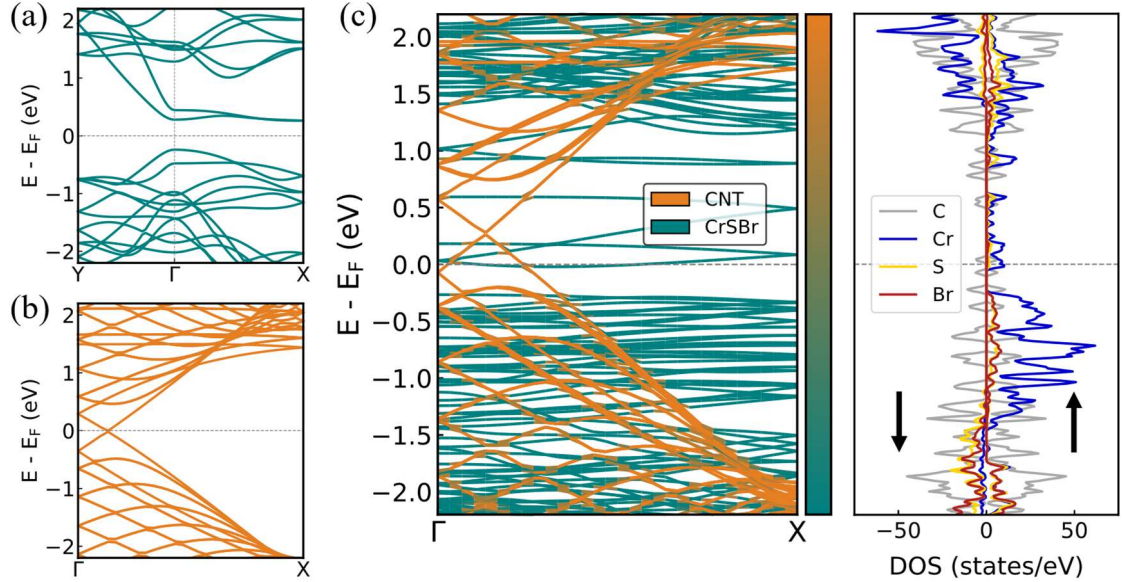

**Figure 2.** Electronic band structure of (a) 2D CrSBr, (b) (16,16) CNT and (c) CrSBr@CNT with atomic contributions and PDOS.

Orbital hybridization at the heterointerface induces charge transfer from the CNT to the CrSBr NR, as shown in Figure 3a. Bader analysis reveals that the maximum charge transfer occurs for CNTs with diameters of ~ 20 – 22 Å, reaching up to 0.034 e/CrSBr formula unit (f.u.) (Figure 3b). These results are consistent with previous DFT studies on related systems, involving NRs encapsulated in CNTs, where a discrete charge transfer occurs from the C atoms to the guest material.[24,34] In contrast, calculations on hybrid heterostructures encapsulating cylindrical chain-like 1D species have reported significantly larger charge transfer ratios,[27,33] likely due to their geometry, which enhances hybridization within the CNTs through closer interfacial interactions. Figure 3c presents the orbital alignment between the heterostructure components, highlighting a noticeable misalignment between CrSBr NR and CNT atoms. This results in discrete charge transfer at the interface (see Section 3 of Supporting Information). The regions of the CNT closest to the NR atoms are primarily responsible for donating electronic density to CrSBr. This is mainly due to the lower potential barrier along the axial direction of the heterostructure (Figures 3d and e).

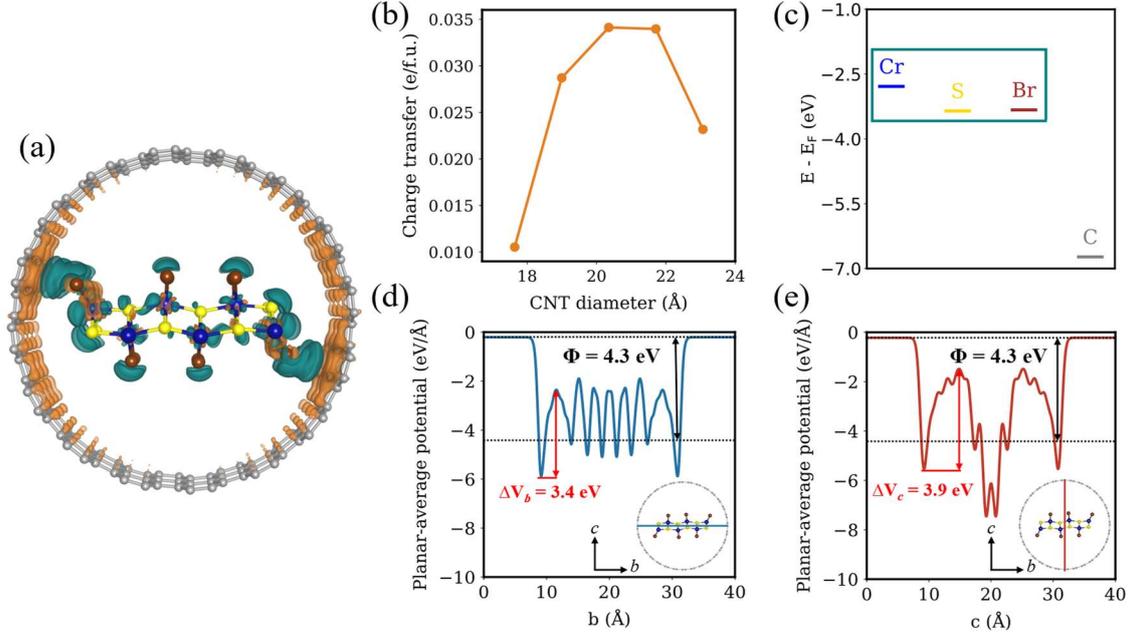

**Figure 3. (a)** Charge density difference after encapsulation of a CrSBr NR within CNT. Color code: orange (cyan) represents charge depletion (accumulation). **(b)** Calculated charge transfer in CrSBr@CNT as a function of CNT diameter. **(c)** Atom-resolved band center of the heterostructure. Planar-averaged electrostatic potential along the **(d)** $b$ and **(e)** $c$ directions. $\Phi$ denotes the work function.

Then, we examine the magnetic properties of CrSBr@CNT, with particular emphasis on how confinement reshapes the magnetic response of the NR. In the CrSBr monolayer, magnetic Cr atoms couple ferromagnetically, primarily through three nearest-neighbor (NN) exchange interactions: $J_1$ (NN along the $a/b$ directions), $J_2$ (NN along $a$), and $J_3$ (NN along $b$). Figure 4a presents a schematic illustration of the Js present in a CrSBr NR. The calculated values for each J are reported in Table 1. Interestingly, upon reduction of the dimensionality to the 1D limit, new magnetic interactions emerge at the NR edges ($J_i'$) due to charge redistribution associated with Cr dangling bonds. The inner exchange couplings of the NR closely resemble those of the 2D counterpart, as the inner atoms experience less structural distortion and reduced charge transfer from the CNT compared to the outer ones. In contrast, the edge exchange interactions are significantly enhanced, a trend that has also been reported for related 2D magnets like $CrI_3$, where edge states play a decisive role in strengthening magnetic couplings.[40,41]

After encapsulation within CNTs, we observe a general enhancement of the exchange interactions in CrSBr, mainly driven by charge transfer from the carbon framework to the NR, while the intrinsic structure of the NR remains preserved in the heterostructure. Notably, edge exchange interactions are more strongly affected than the inner ones, in line with the larger magnetic moments found in edge Cr atoms (Table S3). This enhancement arises because edge-localized magnetic states lie closer to the Fermi level (Figure S5), thereby increasing the sensitivity of edge exchange interactions. To further validate this effect, we simulate the evolution of the Js in pristine 1D CrSBr upon excess of electrons (Figure S6), finding that the observed trend closely mirrors the behavior obtained for the heterostructure. The strong FM exchange interactions play a pivotal role, and are crucial for the robust stabilization of ferromagnetism at the 1D limit in the hybrid CrSBr@CNT system. This robustness is further supported by the large energy separation from competing AFM configurations (Figure 4b), which lie hundreds of meV above the FM ground state (Figure 4c).

Regarding magnetic anisotropy in CrSBr, pristine monolayers exhibit an easy axis of magnetization along the *b* crystallographic direction, an intermediate axis along *a*, and a hard axis along *c*.[42] We calculate magnetocrystalline anisotropy energy (MCE) as: MCE = MAE + MSA, where the magnetic anisotropy energy (MAE) originates from spin–orbit coupling (SOC), and the magnetic shape anisotropy (MSA) arises from dipole–dipole interactions. Together, these terms determine the preferred direction of magnetization in the system. Figure 4d shows the calculated MCE for CrSBr encapsulated within different CNTs, clearly demonstrating that the easy and hard magnetization axes remain along the *b* and *c* directions, respectively. Interestingly, the MCE values in CrSBr@CNT closely match those calculated for the pristine 2D monolayer (11 and 95 µeV/f.u. along the *a* and *c* axes, respectively). These results indicate that CrSBr can be feasibly miniaturized to the 1D limit while preserving its remarkable magnetic properties.

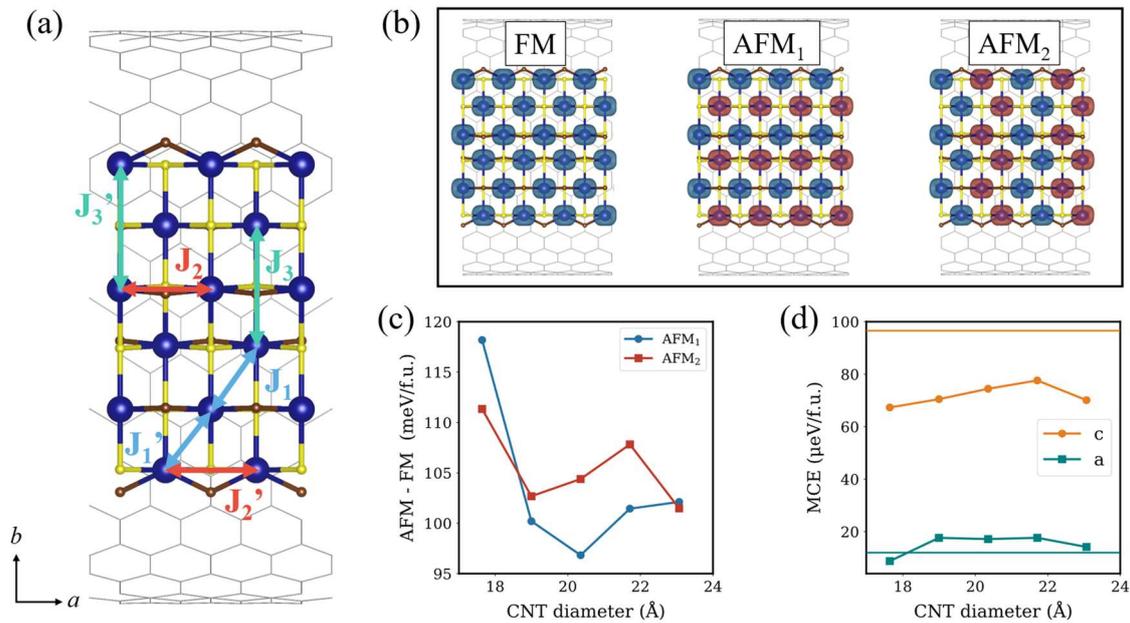

**Figure 4.** (a) Schematic representation of the exchange interactions (J) in 1D CrSBr encapsulated within CNTs. (b) Spin density distribution for the different magnetic configurations considered. Evolution of (c) AFM – FM energy difference and (d) MCE as a function of CNT diameter. Solid lines in (d) indicate the corresponding MCE values calculated for 2D CrSBr, shown as a reference.

**Table 1.** Magnetic exchange couplings (J, in meV) obtained from DFT+U calculations for 2D CrSBr, freestanding 1D CrSBr NR, and encapsulated CrSBr@CNT.

|  | $J_1$ | $J_1'$ | $J_2$ | $J_2'$ | $J_3$ | $J_3'$ |
|---|---|---|---|---|---|---|
| **CrSBr 2D** | 5.08 |  | 3.17 |  | 5.58 |  |
| **CrSBr 1D** | 4.03 | 5.27 | 3.48 | 7.65 | 7.25 | 8.12 |
| **CrSBr 1D@CNT** | 4.00 | 6.04 | 3.65 | 11.27 | 5.83 | 10.21 |

Finally, we conduct spin dynamics simulations using the calculated magnetic parameters of the CrSBr@CNT heterostructure to assess its potential for spintronic applications. Our results reveal a $T_C$ of 50 K (Figure 5a), which is notably lower than that of the CrSBr monolayer. This reduction arises mainly from dimensional confinement, which limits the number of magnetic interactions along the *b* direction. Nevertheless, this value is remarkable when compared with reported transition temperatures of 5 K for $V_xTe_y$ and 3 K for $CrCl_3$ 1D chains encapsulated by CNTs.[23,25] Furthermore, we calculate the magnon dispersion of the CrSBr@CNT hybrid structure using linear spin-wave theory (LSWT)[43] (Figure 5b). The resulting dispersion consists of six bands originating from size quantization along the *b* direction, corresponding to acoustic (A) and optical (O) magnon modes. While the first two acoustic branches ($A_1$ and $A_2$) remain well separated near the Γ point, the third acoustic mode exhibits strong hybridization with the optical magnons (see Figure 5c and Section 6 of the Supporting Information). The acoustic branches closely resemble those observed in the 2D counterpart (Figure S8), thus preserving the distinctive magnetic properties of CrSBr. To examine magnon transport at the macroscopic scale, we simulate the response of extended samples to a localized, short-time magnetic excitation (see in Section 7 in Supporting Information) using the MuMax3 software.[44] Figures 5d–f show the non-equilibrium magnetization distribution 16 ns after the pulse for CrSBr monolayer, NR, and hybrid CrSBr@CNT heterostructure. Upon dimensional reduction, spin waves become confined along *a* direction. Excitation generates magnons with a range of wavelengths, where short wavelength magnons exhibit higher group velocities (Figures S9 and S10). Indeed, magnons in CrSBr@CNT propagate faster than in the pristine CrSBr NR, while both 1D systems exhibit higher group velocities than the 2D CrSBr monolayer along the same direction. Therefore, these results support a confinement-driven emergence of 1D magnetism in CrSBr@CNT.

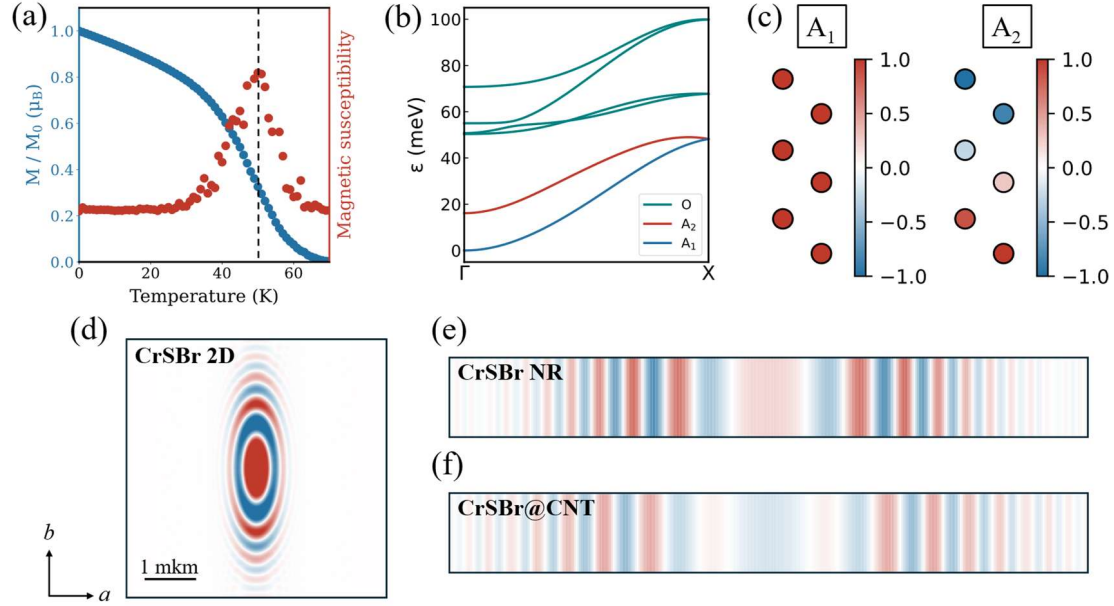

**Figure 5.** (a) Normalized magnetization (blue) and magnetic susceptibility (red) in CrSBr@CNT as a function of temperature. (b) Simulated magnon dispersion in CrSBr@CNT. Color code: 1st (blue) and 2nd (red) acoustic and optical (cyan) magnon modes. (c) Wavefunctions of the 1st ($A_1$) and 2nd ($A_2$) acoustic magnon modes respectively on Cr atoms along $b$ direction. Macroscopic distribution of magnons excited by a local pulse after 16 ns for (d) 2D CrSBr, (e) 1D CrSBr NR and (f) CrSBr@CNT.

In summary, we have investigated the feasibility of achieving 1D magnetism in CrSBr by encapsulation within CNTs through first-principles calculations. We demonstrate that the inner cavity of CNTs can effectively host CrSBr NRs, yielding a hybrid heterostructure with enhanced thermodynamic stability driven by charge transfer from the CNT framework. The pronounced structural and electronic anisotropy of CrSBr enables its confinement and facilitates the preservation of its intrinsic magnetic properties down to the 1D limit. This is confirmed by our magnetic exchange and anisotropy calculations, which unveil that robust ferromagnetic correlations persist upon confinement, closely resembling those of the 2D counterpart. Importantly, the interaction with the CNT enhances magnetic response, leading to a confinement-driven stabilization of ferromagnetism up to 50 K. Furthermore, our spin dynamics simulations predict the propagation of coherent magnons along the confined direction, establishing CrSBr@CNT as a promising platform for exploring 1D magnonics. These findings establish CrSBr as a compelling candidate for 1D magnetism, highlight CNT for engineering quantum-confined magnetic states, and open new avenues for the design of nanoscale spintronic and magnonic devices.

## METHODS

DFT calculations were performed using SIESTA code.[45,46] The generalized gradient approximation (GGA) with Perdew–Burke–Ernzerhof (PBE) parametrization was used to account for exchange-correlation energy[47] and vdW interactions were considered using the Grimme D3 approximation.[48] Atomic coordinates of CrSBr NR were relaxed until forces were less than 0.04 eV Å$^{-1}$ and the positions of carbon atoms in CNTs were kept fixed. A 9 × 1 × 1 $k$-point grid was used to sample reciprocal space for heterostructure calculations. A vacuum layer of 18 Å was introduced in $b$ and $c$ directions to avoid periodic interactions. Double-$\zeta$ polarized basis set was used for all atoms in combination with a real-space mesh cutoff of 900 Ry. AIMD simulations were performed using the NVT canonical ensemble for 5 ps with a time step of 1 fs. Norm-conserving fully relativistic pseudo-potentials taken from the Pseudo-Dojo database in the psml format were used. Charge transfer at the heterointerface was simulated via Bader charge analysis.[49] TB2J package was used to calculate magnetic exchange couplings.[50] Curie temperature calculations were performed via atomistic spin dynamics simulations as implemented in Vampire software,[51] where a 600 mm × 30 nm × 30 nm sample was considered in combination with 10 000 equilibration and loop time steps. Micromagnetic real-space simulations were performed using MuMax3[44] (see Section 7 in Supporting Information).

## NOTES

The authors declare no competing financial interest.

## ACKNOWLEDGEMENTS

The authors acknowledge financial support from the European Union (ERC-2021-StG-101042680 2D-SMARTiES), the María de Maeztu Centre of Excellence Program CEX2024-001467-M funded by MICIU/AEI/10.13039/501100011033, and the Generalitat Valenciana (grant CIDEXG/2023/1). A.M.R. thanks the Spanish MIU (Grant No FPU21/04195). All calculations were performed on the HAWK cluster of the 2D Smart Materials Lab hosted by Servei d'Informàtica of the Universitatde València.

## REFERENCES

(1) Huang, B.; Clark, G.; Navarro-Moratalla, E.; Klein, D. R.; Cheng, R.; Seyler, K. L.; Zhong, D.; Schmidgall, E.; McGuire, M. A.; Cobden, D. H.; Yao, W.; Xiao, D.; Jarillo-Herrero, P.; Xu, X. Layer-Dependent Ferromagnetism in a van Der Waals Crystal down to the Monolayer Limit. *Nature* 2017, *546* (7657), 270–273.

(2) Gong, C.; Li, L.; Li, Z.; Ji, H.; Stern, A.; Xia, Y.; Cao, T.; Bao, W.; Wang, C.; Wang, Y.; Qiu, Z. Q.; Cava, R. J.; Louie, S. G.; Xia, J.; Zhang, X. Discovery of Intrinsic Ferromagnetism in Two-Dimensional van Der Waals Crystals. *Nature* 2017, *546* (7657), 265–269.

(3) Ahn, E. C. 2D Materials for Spintronic Devices. *NPJ 2D Mater Appl* 2020, *4* (1), 17.

(4) Flebus, B.; Grundler, D.; Rana, B.; Otani, Y.; Barsukov, I.; Barman, A.; Gubbiotti, G.; Landeros, P.; Akerman, J.; Ebels, U.; Pirro, P.; Demidov, V. E.; Schultheiss, K.; Csaba, G.; Wang, Q.; Ciubotaru, F.; Nikonov, D. E.; Che, P.; Hertel, R.; Ono, T.; Afanasiev, D.; Mentink, J.; Rasing, T.; Hillebrands, B.; Kusminskiy, S. V.; Zhang, W.; Du, C. R.; Finco, A.; van der Sar, T.; Luo, Y. K.; Shiota, Y.; Sklenar, J.; Yu, T.; Rao, J. The 2024 Magnonics Roadmap. *Journal of Physics: Condensed Matter* 2024, *36* (36), 363501.

(5) Wang, Q. H.; Bedoya-Pinto, A.; Blei, M.; Dismukes, A. H.; Hamo, A.; Jenkins, S.; Koperski, M.; Liu, Y.; Sun, Q.-C.; Telford, E. J.; Kim, H. H.; Augustin, M.; Vool, U.; Yin, J.-X.; Li, L. H.; Falin, A.; Dean, C. R.; Casanova, F.; Evans, R. F. L.; Chshiev, M.; Mishchenko, A.; Petrovic, C.; He, R.; Zhao, L.; Tsen, A. W.; Gerardot, B. D.; Brotons-Gisbert, M.; Guguchia, Z.; Roy, X.; Tongay, S.; Wang, Z.; Hasan, M. Z.; Wrachtrup, J.; Yacoby, A.; Fert, A.; Parkin, S.; Novoselov, K. S.; Dai, P.; Balicas, L.; Santos, E. J. G. The Magnetic Genome of Two-Dimensional van Der Waals Materials. *ACS Nano* 2022, *16* (5), 6960–7079.

(6) Gibertini, M.; Koperski, M.; Morpurgo, A. F.; Novoselov, K. S. Magnetic 2D Materials and Heterostructures. *Nat Nanotechnol* 2019, *14* (5), 408–419.

(7) Geim, A. K.; Grigorieva, I. V. Van Der Waals Heterostructures. *Nature* 2013, *499* (7459), 419–425.

(8) Kizuka, T. Atomic Configuration and Mechanical and Electrical Properties of Stable Gold Wires of Single-Atom Width. *Phys Rev B* 2008, *77* (15), 155401.

(9) Sokolov, A.; Zhang, C.; Tsymbal, E. Y.; Redepenning, J.; Doudin, B. Quantized Magnetoresistance in Atomic-Size Contacts. *Nat Nanotechnol* 2007, *2* (3), 171–175.

(10) Nadj-Perge, S.; Drozdov, I. K.; Li, J.; Chen, H.; Jeon, S.; Seo, J.; MacDonald, A. H.; Bernevig, B. A.; Yazdani, A. Observation of Majorana Fermions in Ferromagnetic Atomic Chains on a Superconductor. *Science (1979)* 2014, *346* (6209), 602–607.

(11) Guo, J.; Xiang, R.; Cheng, T.; Maruyama, S.; Li, Y. One-Dimensional van Der Waals Heterostructures: A Perspective. *ACS Nanoscience Au* 2022, *2* (1), 3–11.

(12) Yang, X.; Zhao, X.; Liu, T.; Yang, F. Precise Synthesis of Carbon Nanotubes and One-Dimensional Hybrids from Templates †. *Chin J Chem* 2021, *39* (6), 1726–1744.

(13) Fu, L.; Shang, C.; Zhou, S.; Guo, Y.; Zhao, J. Transition Metal Halide Nanowires: A Family of One-Dimensional Multifunctional Building Blocks. *Appl Phys Lett* 2022, *120* (2).

(14) Balandin, A. A.; Kargar, F.; Salguero, T. T.; Lake, R. K. One-Dimensional van Der Waals Quantum Materials. *Materials Today* 2022, *55*, 74–91.

(15) Xia, Y.; Yang, P.; Sun, Y.; Wu, Y.; Mayers, B.; Gates, B.; Yin, Y.; Kim, F.; Yan, H. One-Dimensional Nanostructures: Synthesis, Characterization, and Applications. *Advanced Materials* 2003, *15* (5), 353–389.

(16) Li, Y.; Hu, Z.; Guo, Q.; Li, J.; Liu, S.; Xie, X.; Zhang, X.; Kang, L.; Li, Q. Van Der Waals One-Dimensional Atomic Crystal Heterostructure Derived from Carbon Nanotubes. *Chem Soc Rev* 2025, *54* (11), 5619–5656.

(17) Pham, T.; Oh, S.; Stetz, P.; Onishi, S.; Kisielowski, C.; Cohen, M. L.; Zettl, A. Torsional Instability in the Single-Chain Limit of a Transition Metal Trichalcogenide. *Science (1979)* 2018, *361* (6399), 263–266.

(18) Guo, Q.; Wang, X.; Zhao, P.; Zhang, Z.; Geng, L.; Liu, Y.; Teng, Y.; Zhong, Y.; Kang, L. Performance Enhancement of Carbon Nanotube Network Transistors via SbI 3 Inner-Doping in Selected Regions. *Advanced Materials* 2025, *37* (7).

(19) Che, T.; Liu, S.; Wang, Y.; Zhao, P.; Yang, C.; Pan, X.; Ji, H.; Geng, L.; Sun, Q.; Hu, Z.; Li, A.; Zhou, C.; Xu, L.-C.; Zhong, Y.; Tian, D.; Yang, Y.; Kang, L. Interfacial Charge Transfer in One-Dimensional AgBr Encapsulated inside Single-Walled Carbon Nanotube Heterostructures. *ACS Nano* 2024, *18* (47), 32569–32577.

(20) Liu, S.; Teng, Y.; Zhang, Z.; Lai, J.; Hu, Z.; Zhang, W.; Zhang, W.; Zhu, J.; Wang, X.; Li, Y.; Zhao, J.; Zhang, Y.; Qiu, S.; Zhou, W.; Cao, K.; Chen, Q.; Kang, L.; Li, Q. Interlayer Charge Transfer Induced Electrical Behavior Transition in 1D AgI@sSWCNT van Der Waals Heterostructures. *Nano Lett* 2024, *24* (2), 741–747.

(21) He, Q.; Xu, T.; Li, J.; Wang, J.; Jin, C.; Chen, Q.; Gu, X.; Wang, X.; Wei, J.; Duan, H.; Gong, Y. Confined PdMo Ultrafine Nanowires in CNTs for Superior Oxygen Reduction Catalysis. *Adv Energy Mater* 2022, *12* (26).

(22) Townsend, W. J. V.; López-Alcalá, D.; Bird, M. A.; Jordan, J. W.; Rance, G. A.; Biskupek, J.; Kaiser, U.; Baldoví, J. J.; Walsh, D. A.; Johnson, L. R.; Khlobystov, A. N.; Newton, G. N. The Role of Carbon Catalyst Coatings in the Electrochemical Water Splitting Reaction. *Nat Commun* 2025, *16* (1), 4460.

(23) Li, Y.; Li, A.; Li, J.; Tian, H.; Zhang, Z.; Zhu, S.; Zhang, R.; Liu, S.; Cao, K.; Kang, L.; Li, Q. Efficient Synthesis of Highly Crystalline One-Dimensional CrCl 3 Atomic Chains with a Spin Glass State. *ACS Nano* 2023, *17* (20), 20112–20119.

(24) Çaha, I.; Ahmad, A. ul; Boddapatti, L.; Bañobre-López, M.; Costa, A. T.; Enyashin, A. N.; Li, W.; Gargiani, P.; Valvidares, M.; Fernández-Rossier, J.; Deepak, F. L. One-Dimensional CrI3 Encapsulated within Multi-Walled Carbon Nanotubes. *Commun Chem* 2025, *8* (1), 155.

(25) Lan, X.; Geng, L.; Zhang, Z.; Li, Y.; Yuan, J.; Zhou, C.-X.; Huang, S.; Hu, Z.; Li, J.; Yang, C.; Zhang, Y.; Fan, Z.; Tian, D.; Zhao, X.; Li, Q.; Kang, L. Tunable Synthesis of Atomic One-Dimensional VxTey Magnets within Single-Walled Carbon Nanotubes. *Nat Commun* 2025, *16* (1), 6300.

(26) Zhao, X.; Wang, K.; Li, B.; Xiao, Q.; Song, M.; Wang, W.; Zhang, L.; Yao, F.; Yu, B.; Li, Y.; Wang, X.; Guo, S.; Jin, C.; He, J.; Yang, F. Growth of Atomically Thin Metastable β-Tungsten in Single-Walled Carbon Nanotubes for Stable One-Dimensional Ferromagnets. *J Am Chem Soc* 2025, *147* (8), 7028–7038.

(27) Lee, Y.; Choi, Y. W.; Lee, K.; Song, C.; Ercius, P.; Cohen, M. L.; Kim, K.; Zettl, A. 1D Magnetic MX3 Single-Chains (M = Cr, V and X = Cl, Br, I). *Advanced Materials* 2023, *35* (49). https://doi.org/10.1002/adma.202307942.

(28) Lee, Y.; Choi, U.; Kim, K.; Zettl, A. Recent Progress in Realizing Novel One-Dimensional Polymorphs via Nanotube Encapsulation. *Nano Converg* 2024, *11* (1), 52.

(29) Meyer, S.; Pham, T.; Oh, S.; Ercius, P.; Kisielowski, C.; Cohen, M. L.; Zettl, A. Metal-Insulator Transition in Quasi-One-Dimensional HfTe3 in the Few-Chain Limit. *Phys Rev B* 2019, *100* (4), 041403.

(30) Lee, Y.; Choi, Y. W.; Lee, K.; Song, C.; Ercius, P.; Cohen, M. L.; Kim, K.; Zettl, A. Tuning the Sharing Modes and Composition in a Tetrahedral GeX 2 (X = S, Se) System via One-Dimensional Confinement. *ACS Nano* 2023, *17* (9), 8734–8742.

(31) Cain, J. D.; Oh, S.; Azizi, A.; Stonemeyer, S.; Dogan, M.; Thiel, M.; Ercius, P.; Cohen, M. L.; Zettl, A. Ultranarrow TaS 2 Nanoribbons. *Nano Lett* 2021, *21* (7), 3211–3217.

(32) Pham, T.; Oh, S.; Stonemeyer, S.; Shevitski, B.; Cain, J. D.; Song, C.; Ercius, P.; Cohen, M. L.; Zettl, A. Emergence of Topologically Nontrivial Spin-Polarized States in a Segmented Linear Chain. *Phys Rev Lett* 2020, *124* (20), 206403.

(33) Stonemeyer, S.; Cain, J. D.; Oh, S.; Azizi, A.; Elasha, M.; Thiel, M.; Song, C.; Ercius, P.; Cohen, M. L.; Zettl, A. Stabilization of NbTe 3 , VTe 3 , and TiTe 3 via Nanotube Encapsulation. *J Am Chem Soc* 2021, *143* (12), 4563–4568.

(34) Stonemeyer, S.; Dogan, M.; Cain, J. D.; Azizi, A.; Popple, D. C.; Culp, A.; Song, C.; Ercius, P.; Cohen, M. L.; Zettl, A. Targeting One- and Two-Dimensional Ta–Te Structures via Nanotube Encapsulation. *Nano Lett* 2022, *22* (6), 2285–2292.

(35) Wang, Y.; Xiao, J.; Zhu, H.; Li, Y.; Alsaid, Y.; Fong, K. Y.; Zhou, Y.; Wang, S.; Shi, W.; Wang, Y.; Zettl, A.; Reed, E. J.; Zhang, X. Structural Phase Transition in Monolayer MoTe2 Driven by Electrostatic Doping. *Nature* 2017, *550* (7677), 487–491.

(36) Ziebel, M. E.; Feuer, M. L.; Cox, J.; Zhu, X.; Dean, C. R.; Roy, X. CrSBr: An Air-Stable, Two-Dimensional Magnetic Semiconductor. *Nano Lett* 2024, *24* (15), 4319–4329.

(37) Lee, K.; Dismukes, A. H.; Telford, E. J.; Wiscons, R. A.; Wang, J.; Xu, X.; Nuckolls, C.; Dean, C. R.; Roy, X.; Zhu, X. Magnetic Order and Symmetry in the 2D Semiconductor CrSBr. *Nano Lett* 2021, *21* (8), 3511–3517.

(38) Klein, J.; Pingault, B.; Florian, M.; Heißenbüttel, M.-C.; Steinhoff, A.; Song, Z.; Torres, K.; Dirnberger, F.; Curtis, J. B.; Weile, M.; Penn, A.; Deilmann, T.; Dana, R.; Bushati, R.; Quan, J.; Luxa, J.; Sofer, Z.; Alù, A.; Menon, V. M.; Wurstbauer, U.; Rohlfing, M.; Narang, P.; Lončar, M.; Ross, F. M. The Bulk van Der Waals Layered Magnet CrSBr Is a Quasi-1D Material. *ACS Nano* 2023, *17* (6), 5316–5328.

(39) Liu, W.; Guo, X.; Schwartz, J.; Xie, H.; Dhale, N. U.; Sung, S. H.; Kondusamy, A. L. N.; Wang, X.; Zhao, H.; Berman, D.; Hovden, R.; Zhao, L.; Lv, B. A Three-Stage Magnetic Phase Transition Revealed in Ultrahigh-Quality van Der Waals Bulk Magnet CrSBr. *ACS Nano* 2022, *16* (10), 15917–15926.

(40) Qiu, G.; Li, Z.; Zhou, K.; Cai, Y. Flexomagnetic Noncollinear State with a Plumb Line Shape Spin Configuration in Edged Two-Dimensional Magnetic CrI3. *NPJ Quantum Mater* 2023, *8* (1), 15.

(41) Jiang, W.; Li, S.; Liu, H.; Lu, G.; Zheng, F.; Zhang, P. First-Principles Calculations of Magnetic Edge States in Zigzag CrI3 Nanoribbons. *Phys Lett A* 2019, *383* (8), 754–758.

(42) Yang, K.; Wang, G.; Liu, L.; Lu, D.; Wu, H. Triaxial Magnetic Anisotropy in the Two-Dimensional Ferromagnetic Semiconductor CrSBr. *Phys Rev B* 2021, *104* (14), 144416.

(43) Toth, S.; Lake, B. Linear Spin Wave Theory for Single-Q Incommensurate Magnetic Structures. *Journal of Physics: Condensed Matter* 2015, *27* (16), 166002.

(44) Vansteenkiste, A.; Leliaert, J.; Dvornik, M.; Helsen, M.; Garcia-Sanchez, F.; Van Waeyenberge, B. The Design and Verification of MuMax3. *AIP Adv* 2014, *4* (10).

(45) Soler, J. M.; Artacho, E.; Gale, J. D.; García, A.; Junquera, J.; Ordejón, P.; Sánchez-Portal, D. The SIESTA Method for *Ab Initio* Order- *N* Materials Simulation. *Journal of Physics: Condensed Matter* 2002, *14* (11), 2745–2779.

(46) García, A.; Papior, N.; Akhtar, A.; Artacho, E.; Blum, V.; Bosoni, E.; Brandimarte, P.; Brandbyge, M.; Cerdá, J. I.; Corsetti, F.; Cuadrado, R.; Dikan, V.; Ferrer, J.; Gale, J.; García-Fernández, P.; García-Suárez, V. M.; García, S.; Huhs, G.; Illera, S.; Korytár, R.; Koval, P.; Lebedeva, I.; Lin, L.; López-Tarifa, P.; Mayo, S. G.; Mohr, S.; Ordejón, P.; Postnikov, A.; Pouillon, Y.; Pruneda, M.; Robles, R.; Sánchez-Portal, D.; Soler, J. M.; Ullah, R.; Yu, V. W.; Junquera, J. Siesta : Recent Developments and Applications. *J Chem Phys* 2020, *152* (20).

(47) Perdew, J. P.; Burke, K.; Ernzerhof, M. Generalized Gradient Approximation Made Simple. *Phys Rev Lett* 1996, *77* (18), 3865–3868.

(48) Grimme, S.; Antony, J.; Ehrlich, S.; Krieg, H. A Consistent and Accurate *Ab Initio* Parametrization of Density Functional Dispersion Correction (DFT-D) for the 94 Elements H-Pu. *J Chem Phys* 2010, *132* (15), 154104.

(49) Henkelman, G.; Arnaldsson, A.; Jónsson, H. A Fast and Robust Algorithm for Bader Decomposition of Charge Density. *Comput Mater Sci* 2006, *36* (3), 354–360.

(50) He, X.; Helbig, N.; Verstraete, M. J.; Bousquet, E. TB2J: A Python Package for Computing Magnetic Interaction Parameters. *Comput Phys Commun* 2021, *264*, 107938.

(51) Evans, R. F. L.; Fan, W. J.; Chureemart, P.; Ostler, T. A.; Ellis, M. O. A.; Chantrell, R. W. Atomistic Spin Model Simulations of Magnetic Nanomaterials. *Journal of Physics: Condensed Matter* 2014, *26* (10), 103202.